# Effect of disorder in the transition from topological insulator to valley-spin polarized state in silicene and Germanene

Partha Goswami

Deshbandhu College, University of Delhi, Kalkaji, New Delhi-110019, India Tel:+91-9810163784

physicsgoswami@gmail.com

**ABSTRACT**

We start with the silicene/germanene single-particle Hamiltonian in buckled 2D hexagonal lattices expressed in terms of $4\times 4$ Dirac matrices($\gamma^{\mu}$) in the Weyl basis. The Hamiltonian of these systems comprises of the Dirac kinetic energy, a mass gap term, and the spin-orbit coupling. The second term breaks the sub- lattice symmetry of the silicene's honey-comb structure and generates a gap. The buckled structure generates a staggered sub-lattice potential between silicon atoms at A sites and B sites for an applied electric field $E_z$ perpendicular to its plane. Tuning of $E_z$, allows for rich behavior varying from a topological insulator (TI)to a band insulator (BI) with a valley spin-polarized metal (VSPM) at a critical value in between. Thus, the mobile electrons in silicene/germanene are coupled differently, compared to graphene, to an external (tunable)electric field. Our preliminary investigation have shown that, as long as the (non-magnetic) impurity scattering strength $V_0$ is moderate , i.e. $V_0$ is of the same order as the intrinsic spin-orbit coupling $t_{so}$(~ 4 meV), VSPM phase is protected. The effective "two-component Dirac physics" remains valid in this phase. The increase in $E_z$ beyond the critical value leads to the valley magnetic moment reversal. The enhancement in $V_0$, however, leads to the disappearance of the VSPM phase due to accentuated intra- and inter-valley scattering processes. This disappearance does not occur due to the increase in Rashba spin-orbit coupling effect .

## 1. INTRODUCTION

A single layer of silicon atoms (and possibly the only crystalline allotrope of silicon), called silicene, has been synthesized on the Ag(111) substrate recently exhibiting an analogous honeycomb structure as graphene [1,2,3,4,5]. The silicene sheet, in fact, has linearly crossing bands at the **K** and **K′** symmetry points. Thus, the charge carriers in silicene behave like relativistic particles with a conical energy spectrum(and Fermi velocity $v_F \approx 10^6$ m-s$^{-1}$), as in graphene. There had also been silicene synthesis report by depositing Si atoms on surfaces of Ir [6]. This discovery has given a big boost to the search for materials that host topological insulator (TI) phases [7,8]. The two dimensional TIs exhibit the quantum spin Hall effect (QSHE) with gapless edge states and a finite energy gap in the bulk [9,10]. The first proposal of this exotic state of matter was made by Kane and Mele [9] considering graphene in the presence of spin-orbit coupling(SOC). But the insignificant SOC in graphene rendered the QSHS inaccessible in that material. However, since silicon is heavier than carbon, the spin-orbit coupling in silicene is naturally much larger than in graphene. It is thus feasible to experimentally access QSHE in silicene. The unit cell of silicene contains two atoms which gives rise to two different sub-lattices A and B as in graphene. The honeycomb lattice of the former system, however, is distorted due to a large ionic radius of a silicon atom and forms a buckled structure pointing out-of-plane [11]. Furthermore, the stronger SOC in silicene has its origin also in the buckled structure of the former. The A and B sites here form two sub-lattices separated by a perpendicular distance, say, $2\ell$. The structure generates a staggered sub-lattice potential $2\ell E_z$ between silicon atoms at A sites and B sites for an applied electric field $E_z$ perpendicular to the silicene-plane. Thus, the mobile electrons in silicene are able to couple differently to an external electric field than the ones in graphene. This difference is the origin of new Rashba spin-orbit coupling effects that allow for external tuning and closing of the band gap in silicene [12]. The tunable band-gap facilitates silicene's potential use in micro-electronics. Due to the interplay of spin-orbit coupling and the electric field strength, the silicene bands display spin-valley locking:

Tuning of $E_z$, allows for rich behavior varying from a topological insulator to a band insulator with a valley spin-polarized metal at a critical value ($E_c$ ) in between. In fact, as in refs.[13,14,15], at the critical point with ($E_z/E_c$ ) = 1.00 the gap of one of the spin-split bands closes to give a Dirac point while at the other **K** point it is gapped. Furthermore, it is the other spin band which has no gap. This spin-valley locked phase has been termed a valley-spin-polarized metal (VSPM). This topological phase transition can be detected experimentally by way of diamagnetism [16]. The emergence of single Dirac cone state (where the gap is open at the **K** point and closed at the **K'** point [17])is also possible when we apply photo-irradiation and electric field.

The purpose of this communication is to report the investigation of stand-alone low-buckled monolayer silicene(MLS), assuming that the Rashba SOC which includes spin- flip processes to be present and, focusing on the the most intriguing property of the silicene, viz. the opposite spin polarization at different valleys, i.e., the valley-spin locking. Explicitly, the Dirac cone around **K** (**K′** ) point is polarized with spin up (down), mainly originating from the intrinsic SOC between next nearest-neighbor (NNN) sites as well as broken inversion symmetry due to the external electric field. The Rashba SOC also acting between NNN sites is in the nature of the correction to the intrinsic SOC effects. Ideally, the spin around each cone is fully polarized here, and the spin-flip and the inter-valley scattering from non-magnetic impurities is strictly prohibited by time reversal symmetry (TRS). Therefore, two Dirac cones in this system are effectively decoupled and consequently the two-component, single-flavor Dirac physics emerges. Now it is quite imperative to ask (i) if there can be any delocalized states in the strict sense under disorder, and (ii) can the Rashba SOC, inevitable in realistic silicene, induce inter-valley scattering and lead to the breakdown of the single Dirac cone physics as well? We find answers to these questions here. We conclude that the stability of the SVPM phase is not topologically protected against dissipation and fluctuation in the presence of large defects.

The paper is organized as follows: In Sec. II, a brief outline of the tight binding model of silicene is given and the low-energy excitation spectrum is obtained. In Sec.III, the Born scattering approximation followed by the t-matrix approximation to deal with the impurity problem are discussed. The renormalized single-particle excitation spectrum for the finite chemical potential is obtained. The concluding remarks could be found in Sec.IV.

## 2. SINGLE PARTICLE EXCITATION SPECTRUM

### A. The Tight Binding Model

The tight binding model [14,15] describing the silicene system involves six terms. Apart from the usual nearest-neighbor hopping term($-t\sum_{ij,\sigma} c^{\dagger}_{i\sigma} c_{j\sigma}$) with i and j referring to the nearest neighbour sites labeled A and B on the sub-lattices A and B, respectively, $c_{i\sigma}$ is π-orbital annihilation operator for an electron with spin σ on site i ($c^{\dagger}_{i\sigma}$ may also be termed as the quantum amplitudes for an electron to occupy sites labeled i on the sub-lattices A and B), and the transfer energy **t = 1.6 eV,** the effective spin-orbit coupling(SOC) term which, in coordinate representation, may be written as $H_{so} = (it_{so}/3\sqrt{3}) \sum_{\langle ij \rangle,\alpha\beta} \nu_{ij} c^{\dagger}_{i\alpha}\sigma^{z}_{\alpha\beta} c_{j\beta}$ where $\langle ij \rangle$ run over all the next-nearest-neighbor hopping sites, **$t_{so}$ =3.9 meV** is the effective SO coupling, $\boldsymbol{\sigma} = (\sigma_x, \sigma_y, \sigma_z)$ is the Pauli matrix of spin, $\nu_{ij} = +1$ if the next-nearest-neighboring hopping is anticlockwise and $\nu_{ij} = -1$ if it is clockwise with respect to the positive z axis. As regards Rashba spin-orbit coupling(RSOC) we have two terms. The first term [$\xi i\, t_1(E_z)\sum_{ij,\alpha\beta} c^{\dagger}_{i\alpha}(\boldsymbol{\sigma} \times \mathbf{d}_{ij})^{z}_{\alpha\beta} c_{j\beta}$] represents the RSOC associated with the nearest neighbor hopping induced by external electric field $E_z$ [5,6,7] where the unit vector $\mathbf{d}_{ij}$ connects two sites **i** and **j** in the same sub-lattice (In fact, the vector $\mathbf{d}_{ij}$ connects the two nearest bonds connecting the next-nearest neighbours. Thus, the two sites **i** and **j** are on the same sub-lattice.). The parameter $t_1(E_z)$ satisfies $t_1(E_z= 0) = 0$ and becomes of the order of 10 µeV at the critical electric field $\mathbf{E_c = t_{so}/\ell} = 17$ meV A$^{O-1}$. The second term represents the second RSOC $[-(\frac{2}{3}) \xi i\, t_2 \sum_{\langle ij \rangle,\alpha\beta} \mu_i c^\dagger_{i\alpha} (\boldsymbol{\sigma} \times \mathbf{d}_{ij})^z_{\alpha\beta} c_{j\beta}]$ with $\mathbf{t_2 \sim 0.7}$ **meV** and the unit vector $\mathbf{d}_{ij}$ connects two sites i and j in the same sub-lattice. The term $\mu_i$ is +1 if 'i' corresponds to A sub-lattice and – 1 if it corresponds to B sub-lattice. The $t_1(E_z)$ term being much smaller than the other terms, we ignore it all together. The term $(-\ell\, E_z \sum_{i,\sigma} \mu_i c^\dagger_{i\sigma} c_{i\sigma})$ with $\boldsymbol{\ell} =$ **0.23 Å** is the staggered sublattice potential term where once again $\mu_i = \pm 1$ for the A(B) site. These terms break the sub-lattice symmetry of the silicene's honey-comb structure and generate a gap. Opening a gap in graphene by these means is not possible as the A and B sub-lattices lie in the same plane. Apart from these terms, there may be a term involving the exchange field M. This term may be written as $[(M \sum_{i,\sigma} c^\dagger_{i\sigma} \sigma_z c_{i\sigma})]$. The exchange field M may arise due to proximity coupling to a ferromagnet such as depositing Fe atoms to the silicene surface or depositing silicene to a ferromagnetic insulating substrate. The model Hamiltonian can also be used to describe germanene, which is a honeycomb structure of germanium[5, 6], where various parameters are t = 1.3eV, $\mathbf{t_{so}}$ = 43meV, $t_2$ = 10. 7 meV and $\ell$ = 0.33Å.

**B. The Low-energy Excitation Spectrum**

By performing Fourier transformations, one obtains the low-energy effective Hamitonian around Dirac points **K** and **K′**, say, in the basis $c_{\delta \mathbf{k},\beta} = (a_{\delta \mathbf{k} \uparrow},\ b_{\delta \mathbf{k}\uparrow},\ a_{\delta \mathbf{k}\downarrow},\ b_{\delta \mathbf{k}\downarrow})$ in momentum space. We calculate the electronic band dispersion of silicene around these points. Here $a_{\delta \mathbf{k},\sigma}$ and $b_{\delta \mathbf{k},\sigma}$

(with σ =↑↓) correspond to the fermion annihilation operators for the single-particle state (**k**,σ). The single-particle low-energy Hamiltonian may be written in a compact form in terms of Dirac matrices($\gamma^{\mu}$) in the Weyl framework as $H = \sum_{\delta \mathbf{k},\alpha,\beta} c^{\dagger}_{\delta \mathbf{k},\alpha} \hbar(\delta \mathbf{k}) c^{\dagger}_{\delta \mathbf{k},\beta}$, where the matrix

$$\hbar(\delta \mathbf{k})/\left(\frac{\hbar v_F}{a}\right) = [\xi \, a \, v^x \, \delta k_x + a \, v^y \, \delta k_y] + \xi \, [\Delta_z \times (\gamma^5 \gamma^z \gamma^0) + t'_{so} \times (\gamma^5 \gamma^z \gamma^0 \gamma^5)]$$

$$+ \xi \, [a \, i \, t'_2 \, (\gamma^z \delta k_x + i\gamma^5 \gamma^z \delta k_y) - (M / \left(\frac{\hbar v_F}{a}\right))(\gamma^5 \gamma^z \gamma^0)] , \quad (1)$$

$$\Delta_z = \ell E'_z = \frac{\ell E_z}{\left(\frac{\hbar v_F}{a}\right)}, \Delta_{so} = t'_{so} = \frac{t_{so}}{\left(\frac{\hbar v_F}{a}\right)}, v_x = (\gamma^5 \gamma^0 \gamma^x), v_y = (\gamma^5 \gamma^0 \gamma^y), \quad (2)$$

In the absence of the Rashba and the exchange terms the Hamiltonian appears as $^{sz}H_{\xi}/\left(\frac{\hbar v_F}{a}\right) = [-\xi \, a \, \delta k_x \, \sigma_x + a \, \delta k_y \, \sigma_y] + \xi s_z \Delta_{so} \sigma_z + \Delta_z \sigma_z$. The Hamiltonian basically comprises of the kinetic energy term involving the velocity operators $v_x$ and $v_y$, the sub-lattice symmetry (of the silicene's honey-comb structure) breaking term generating a mass-gap, and the spin-orbit coupling. Opening a gap in graphene by these means is not possible as the A and B sub-lattices lie in the same plane. The 4×4 matrices are given by $\gamma^0 = \begin{pmatrix} 0 & I_2 \\ I_2 & 0 \end{pmatrix}, \gamma^i = \begin{pmatrix} 0 & \sigma_i \\ -\sigma_i & 0 \end{pmatrix}, \gamma^5 = \begin{pmatrix} -I_2 & 0 \\ 0 & I_2 \end{pmatrix}$, $I_2$ denotes the 2 × 2 identity matrix, $\sigma_i$ denote the Pauli matrices. The terms involving the second Rashba SOC $t_2$ may also go as part of the velocity operators. The eigenvalues($\epsilon(\mathbf{k})$) of $\hbar_K(\mathbf{k})$ above are given by the equation $\det(\hbar(\delta \mathbf{k}) - \epsilon(\delta \mathbf{k}) I_4 \times I_4) = 0$. Assuming that proximity coupling to a ferromagnet is not accessible, we find that the spin-split bands close to a Dirac point in the absence of the exchange field are given by $\frac{\epsilon(\delta \mathbf{k})_{\xi,s_z}}{\left(\frac{\hbar v_F}{a}\right)} \approx \pm [(a|\delta \mathbf{k}|)^2 + \{\Delta_{soc}(a|\delta \mathbf{k}|) + \xi s_z \Delta_z\}^2]^{1/2}$. One observes that the effect of the applied electric field $E_z$ is to lift the spin-valley degeneracy. We put total spin-orbit coupling gap as

$(t'^2_{so}+(at'_2|\delta \mathbf{k}|)^2)^{1/2}=(a^2/(D\tau_{so})+(at'_2|\delta \mathbf{k}|)^2)^{1/2}= \Delta_{SOC}$ $|\delta \mathbf{k}|)$ and $s_z = \pm 1$ for $\{\uparrow, \downarrow\}$. We note that the intrinsic and the extrinsic consequences of spin-orbit interaction are the Dirac model and the spin-orbit scattering, respectively. We denote by $\ell_{so}$ and $\tau_{so}$ the spin-orbit scattering length and time, respectively. Whereas $\tau_{so}^{-1}$ depends upon the impurity potential for the spin-orbit scattering, a similar quantity $\tau_e^{-1}$ depends upon the impurity potential for the elastic scattering to be introduced shortly. The shorter $\ell_{so}$ ($\ell_{so} = \sqrt{(D\tau_{so})}$, where D is some diffusion constant) means stronger spin-orbit scattering. The time-reversal symmetry is preserved here as $\varepsilon(\xi,s_z,a\delta\mathbf{k}) = \varepsilon(\xi,s_z,-a\delta\mathbf{k})$.

In the absence of the exchange field and the applied electric field $E_z$, one finds that the intrinsic SOC and the Rashba SOC terms effective between the same sub-lattice are in quadrature with the leading hopping term ($\hbar v_F |\delta\mathbf{k}|$) in the single-particle excitation spectrum(SPES). In the presence of the applied electric field, the corresponding gap term together with the SOC terms will be in quadrature with the leading hopping term. Thus, the effect of the intrinsic and the Rashba SOC together with the electric field is to impart fermions with mass as they correspond to an effective staggered sub-lattice potential $V(\xi,s_z,a|\delta\mathbf{k}|) = \{\Delta_{SOC}(a|\delta\mathbf{k}|) + \xi s_z \Delta_z\}$.

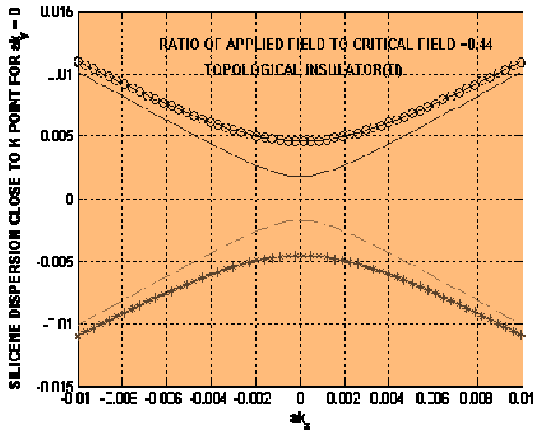
(a)

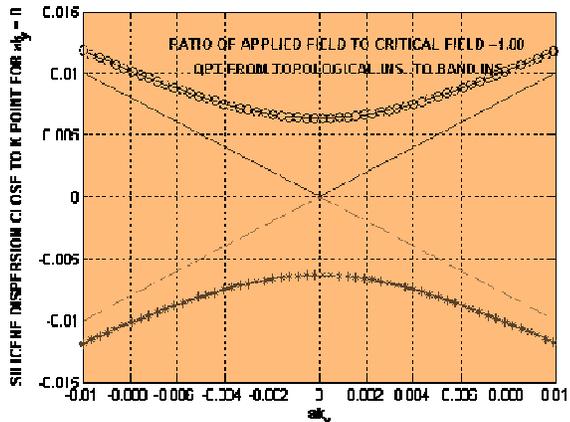
(b)

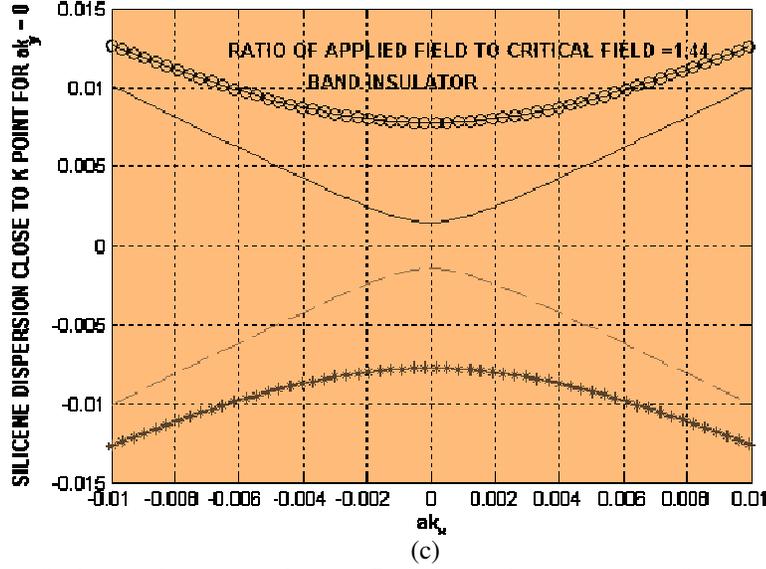

(c)

**Figure1.** A plot of silicene dispersion close to Dirac neutrality point as a function of ($k_x$a) for ($k_y$a) = 0. The presence of spin orbit coupling and a perpendicular electric field gives rise to spin-split bands about the K point, with two gaps as shown in the upper left panel (a), one of which may be tuned to zero at ($E_z/E_c$) = 1.00 as shown in the right panel (b). The bands at K′ are reversed from those at K. In (a) The ratio of the applied electric field to the critical field($E_z/E_c$) is 0.44. In (b), which corresponds to a critical point with ($E_z/E_c$) = 1.00, the gap of one of the spin-split bands closes to give a Dirac point while at the other K point it is gapped and it is the other spin band which has no gap. This has been termed a valley-spin-polarized metal (VSPM). For ($E_z/E_c$) >1 (for example, the case(c) where the ratio ($E_z/E_c$) is 1.44), the spectrum becomes fully gapped again but the system is a band insulator albeit with unusual chiral properties [18].

A plot of low-energy silicene dispersion close to Dirac neutrality point as a function of ($k_x$a) for ($k_y$a) = 0 is given in Figure 1. Corresponding to the critical point with ($E_z/E_c$) = 1.00, indeed, the gap of one of the spin-split bands closes to give a Dirac point while at the other K point it is gapped and it is the other spin band which has no gap. This has been termed a valley-spin-polarized metal (VSPM). In view of the low-energy spectrum given above, the reduced, massive Dirac model Hamiltonian matrix for the silicene reads $\hbar_{reduced}(\xi,s_z,a\delta\mathbf{k})/(\hbar v_F/a)$ [$\xi a\sigma^x\delta k_x + a\sigma^y\delta k_y + V(\xi,s_z,a|\delta\mathbf{k}|)\sigma^z - (\mu/(\hbar v_F/a))\sigma^0$]. Here $\mu' = (\mu/(\hbar v_F/a))$ is the dimensionless

chemical potential of the fermion number. The model will be used below to describe the two-dimensional bulk and surface bands in the thin-film limit in silicene and germanene by introducing different model parameters.

## 3. T-MATRIX APPROXIMATION

The impurity potentials are given by $U(\mathbf{r}) = U_0(\mathbf{r}) + U_{so}(\mathbf{r})$, where $U_0(\mathbf{r}) = \sum_i u_0(\mathbf{r}-\mathbf{R}_i)$ is for the elastic scattering, and $U_{so}(\mathbf{r}) = \sum_i (\hbar/4m^2c^2) \boldsymbol{\sigma} \cdot (\nabla u_{so}(\mathbf{r} - \mathbf{R}_i) \times \boldsymbol{p})$ is for the spin-orbit scattering. We assume that all non-magnetic impurities are alike, distributed randomly, and each of them contribute a potential term $u_0(\mathbf{r}-\mathbf{R}_i) = u_0^i \delta(\mathbf{r} - \mathbf{R}_i)$ where $u_0^i$ is the potential due to a single impurity at $\mathbf{R}_i$. The potential $U_0(\mathbf{r})$ may now be expanded as $U_0(\mathbf{r}) = \sum_{q,i} u_0^i \exp[i\mathbf{q}.(\mathbf{r} - \mathbf{R}_i)]$. We shall similarly assume that $u_{so}(\mathbf{r} - \mathbf{R}_i) = (4m^2c^2/\hbar^2) u_{so}^i \delta(\mathbf{r} - \mathbf{R}_i)$ and, therefore, $(\hbar/4m^2c^2) \nabla u_{so}(\mathbf{r} - \mathbf{R}_i) = (\hbar^{-1}) \sum_q u_{so}^i \nabla \exp[i\mathbf{q}.(\mathbf{r} - \mathbf{R}_i)]$ where $u_{so}^i$ is the strength for the spin-orbit scattering. The additivity of the impurity potentials imply that the total scattering time $\tau$ is given by $\tau^{-1} = \tau_{so}^{-1} + \tau_e^{-1}$. Our aim is to calculate the Born scattering amplitudes corresponding to $U_0(\mathbf{r})$ and $U_{so}(\mathbf{r})$ using the eigenvectors of the Hamiltonian matrix. To do this, we proceed with the fact the massive Dirac Hamiltonian matrix $\hbar_{\text{reduced}}(\xi, s_z, \delta\mathbf{k})/\left(\frac{\hbar v_F}{a}\right)$ above describes a conduction band and a valence band. As in ref.[4], the angle-resolved photoelectron spectroscopy (ARPES) experiment has shown that the Dirac point was measured to be 0.3 eV below the Fermi level $E_F$. We may therefore suppose that $E_F$ intersects with only the conduction band, and in the limit of weak scattering $\hbar/\tau \ll E_F$, the valence band becomes irrelevant for transport. We, however, treat the dispersion of both the conduction and the valence bands $\frac{\epsilon(\delta\mathbf{k})_{\xi,s_z}}{\left(\frac{\hbar v_F}{a}\right)} \approx [\pm\{(a|\delta\mathbf{k}|)^2 + (V(\xi, s_z, a|\delta\mathbf{k}|))^2\}^{1/2} - \mu']$ to be the relevant dispersion. The eigenfunction around $\mathbf{K}$ and $\mathbf{K}'$ are, respectively, given by

$$|k\rangle_{K,\pm} = (1/\sqrt{2}) \begin{pmatrix} \exp(-\frac{i\theta_k}{2}) \\ \pm \exp(\frac{i\theta_k}{2})(\sqrt{(v_{1,k}^2+1)} \mp v_{1,k}) \end{pmatrix},$$

$$|k\rangle_{K',\pm} = (1/\sqrt{2}) \begin{pmatrix} \exp(\frac{i\theta_k}{2}) \\ \mp \exp(-\frac{i\theta_k}{2})(\sqrt{(v_{2,k}^2+1)} \mp v_{2,k}) \end{pmatrix}. \quad (3)$$

where, writing **k** in place of momenta δ**k**, the functions $v_{1,k} \equiv (V(\xi=+1, s_z, a|k|)/(\hbar v_F |k|))$, $v_{2,k} \equiv (V(\xi=-1, s_z, a|k|)/(\hbar v_F |k|))$, $V(\xi, s_z, a|k|) = \{\Delta_{soc}(a|k|) + \xi s_z \Delta_z\}$, $\cos(\theta_k) = k_x/|k|$, $\sin(\theta_k) = k_y/|k|$, and $\theta_k = \arctan(k_y/k_x)$. For the real space, the eigenvectors may be written as $S^{-1/4} |k\rangle_K \exp(i\mathbf{k}\cdot\mathbf{r})$ and $S^{-1/4} |k\rangle_{K'} \exp(i\mathbf{k}\cdot\mathbf{r})$ where S is the area of the sheet. The Born scattering amplitude $U_{k,k'}$ now may be expressed as $U_{k,k'} = U^{elastic}_{k,k'} + U^{so}_{k,k'}$, and in terms of concentrations of nonmagnetic($n_0$), and spin-orbit impurities($n_{so}$), one may write

$$U^{elastic}_{k,k'} = S^{-1/2} \int d\mathbf{r}\, [_{K,\pm}\langle k | U_0(\mathbf{r}) | k'\rangle_{K,\pm} + {}_{K',\pm}\langle k | U_0(\mathbf{r}) | k'\rangle_{K',\pm}] e^{i(\mathbf{k'}-\mathbf{k})\cdot\mathbf{r}},$$

$$= S^{-1/2} \int d\mathbf{r} \sum_{q,j} u_0^j e^{i\mathbf{q}\cdot(-\mathbf{R}_j)} [_{K,+}\langle k | k'\rangle_{K,+} + {}_{K,-}\langle k | k'\rangle_{K,-}$$

$$+ {}_{K',+}\langle k | k'\rangle_{K',+} + {}_{K',-}\langle k | k'\rangle_{K',-}] e^{i(\mathbf{q}+\mathbf{k'}-\mathbf{k})\cdot\mathbf{r}}$$

$$= S^{-1/2} \sum_j u_0^j e^{i(\mathbf{k'}-\mathbf{k})\cdot\mathbf{R}_j} [2\cos((\theta_k-\theta_{k'})/2) + \zeta_{1,k}\exp(-i(\theta_k-\theta_{k'})/2)$$

$$+ \zeta_{2,k}\exp(i(\theta_k-\theta_{k'})/2)]. \quad (4)$$

We may write the entire right-hand-side equal to ($n_0^{1/2} V_0(\mathbf{k}-\mathbf{k'})$) where $V_0(\mathbf{k}-\mathbf{k'})$ is complex.

Here $\zeta_{1,k} = [(v_{1,k}^2 + 1)^{1/2}(v_{1,k'}^2 + 1)^{1/2} + v_{1,k}v_{1,k'}]$, and $\zeta_{2,k} = [(v_{2,k}^2 + 1)^{1/2}(v_{2,k'}^2 + 1)^{1/2} + v_{2,k}v_{2,k'}]$. One may consider the spin-orbit interaction in a similar manner:

$$U^{so}_{k,k'} = S^{-1/2} \int d\mathbf{r} \, [_{K,\pm}\langle \mathbf{k} | U_{so}(\mathbf{r}) | \mathbf{k}' \rangle_{K,\pm} + _{K',\pm}\langle \mathbf{k} | U_{so}(\mathbf{r}) | \mathbf{k}' \rangle_{K',\pm}] e^{i(\mathbf{k}'-\mathbf{k})\cdot\mathbf{r}}$$

$$= S^{-1/2} \int d\mathbf{r} \sum_{j,q} i\, u^j_{so} e^{i\mathbf{q}\cdot(-\mathbf{R}_j)} [_{K,\pm}\langle \mathbf{k} | \sigma\cdot(\mathbf{q}\times\mathbf{k}') | \mathbf{k}' \rangle_{K,\pm}$$
$$+ _{K',\pm}\langle \mathbf{k} | \sigma\cdot(\mathbf{q}\times\mathbf{k}') | \mathbf{k}' \rangle_{K',\pm}] e^{i(\mathbf{q}+\mathbf{k}'-\mathbf{k})\cdot\mathbf{r}}$$

$$= S^{-1/2} \sum_j i\, u^j_{so} e^{i(\mathbf{k}'-\mathbf{k})\cdot\mathbf{R}_j} [_{K,\pm}\langle \mathbf{k} | \sigma\cdot(\mathbf{k}\times\mathbf{k}') | \mathbf{k}' \rangle_{K,\pm}$$
$$+ _{K',\pm}\langle \mathbf{k} | \sigma\cdot(\mathbf{k}\times\mathbf{k}') | \mathbf{k}' \rangle_{K',\pm}]. \quad (5)$$

The quantity $\sigma\cdot(\mathbf{k}\times\mathbf{k}')$ is equal to $\sigma^z(k_x k'_y - k_y k'_x)$ and, therefore, the entire right-hand-side may eventually be written as $(n_{so}^{1/2} V_{so}(\mathbf{k}-\mathbf{k}'))$ where $V_{so}(\mathbf{k}-\mathbf{k}')$ is complex.

The effect of elastic scattering by non-magnetic impurities and spin-orbit scattering involve the calculation of the total self-energy $\Sigma(\mathbf{k},\omega_n) = \Sigma_e(\mathbf{k},\omega_n) + \Sigma_{so}(\mathbf{k},\omega_n)$ in terms of the Matsubara frequencies $\omega_n$, which alters the single-particle excitation spectrum in a fundamental way. In the Green's function matrix $\check{G}(\mathbf{k},\omega_n)$ we insert the self-energy by the Dyson's equation $(\check{G}(\mathbf{k},\omega_n))^{-1} = (\check{G}_0(\mathbf{k},\omega_n))^{-1} - \Sigma(\mathbf{k},\omega_n) I_{2\times 2}$, where $I_{2\times 2}$ is the $2\times 2$ unit matrix. The full Green's functions are now given by $G^{(Full)}_{A,A}(\mathbf{k},\omega_n) = G_{A,A}(\mathbf{k},\omega_n)/[1 - G_{A,A}(\mathbf{k},\omega_n)\Sigma(\mathbf{k},\omega_n)]$, and $G^{(Full)}_{B,B}(\mathbf{k},\omega_n) = G_{B,B}(\mathbf{k},\omega_n)/[1 - G_{B,B}(\mathbf{k},\omega_n)\Sigma(\mathbf{k},\omega_n)]$ where the letters A and B refer to the sub-lattices which do not lie in the same plane. However, $G^{(Full)}_{A,B}(\mathbf{k},\omega_n) = G_{A,B}(\mathbf{k},\omega_n)$. We first consider only the contribution of the Fig.2(a). Assuming the elastic scattering by impurities weak, we may write it as $\Sigma_e^{(1)}(\mathbf{k},\omega_n) = n_0 \sum_{k'} |V_0(\mathbf{k}-\mathbf{k}')|^2 G_{\alpha,\alpha}(\mathbf{k}',\omega_n) = \Sigma_{e,0} + \Sigma_{e,0}^{(1)}(\mathbf{k})$ where $\Sigma_{e,0}$ is

shown to be zero below and the function $\Sigma_{e,0}^{(1)}(\mathbf{k})$ is the first order contribution which can be shown to be independent of $\omega_n$, i.e.

$$\Sigma_{e,0}^{(1)}(\mathbf{k}) = -n_0 \sum_{\mathbf{k}'} |V_0(\mathbf{k}-\mathbf{k}')|^2 (i\omega_n) \int_{-\infty}^{+\infty} d\varepsilon\, \rho(\varepsilon)$$
$$\times [\, u_\mathbf{k}^2\, (\omega_n^2 + E_\mathbf{k}^{(U)\,2})^{-1} + v_\mathbf{k}^2\, (\omega_n^2 + E_\mathbf{k}^{(L)2})^{-1}\,], \qquad (6)$$

$E_\mathbf{k}^{((U),(L))} = \pm\{(a|\mathbf{k}|)^2 + (V(\xi, s_z, a|\mathbf{k}|))^2\}^{1/2} - \mu$, $(u_\mathbf{k}^2, v_\mathbf{k}^2) = (1/2)[1 \mp (V/\{(a|\mathbf{k}|)^2 + V^2\}^{1/2})]$, and $\rho(\varepsilon) = \rho_0[\delta(\varepsilon - E_\mathbf{k}^{(U)}) + \delta(\varepsilon - E_\mathbf{k}^{(L)})]$. We thus obtain $G_{\alpha,\alpha}(\mathbf{k}',\omega_n) \approx -\rho_0 (i\omega_n)(\pi/|\omega_n|)$, and $\Sigma_{e,0}^{(1)}(\mathbf{k}) = -n_0\rho_0 (i\omega_n)\sum_{\mathbf{k}'} |V_0(\mathbf{k}-\mathbf{k}')|^2 (\pi/|\omega_n|) = [-i\omega_n/(2|\omega_n|\tau_{\mathbf{k},e})]$, where $\tau_{\mathbf{k},e}^{-1} = 2\pi n_0 \rho_0 \sum_{\mathbf{k}'} |V_0(\mathbf{k}-\mathbf{k}')|^2$. Similarly, $\Sigma_{so,0}^{(1)}(\mathbf{k}) = -n_{so}\rho_0(i\omega_n)\sum_{\mathbf{k}'} |V_{so}(\mathbf{k}-\mathbf{k}')|^2 (\pi/|\omega_n|) = [-i\omega_n/(2|\omega_n|\tau_{\mathbf{k},so})]$, where $\tau_{\mathbf{k},so}^{-1} = 2\pi n_{so}\rho_0 \sum_{\mathbf{k}'} |V_{so}(\mathbf{k}-\mathbf{k}')|^2$. The total first order self-energy contribution $\Sigma^{(1)}(\mathbf{k})$ independent of magnitude of $\omega_n$ is, thus, $[-i\omega_n/(2|\omega_n|\tau_\mathbf{k})]$ where $\tau_\mathbf{k}^{-1} = \tau_{\mathbf{k},e}^{-1} + \tau_{\mathbf{k},so}^{-1}$. Note that $\tau_\mathbf{k}$, which corresponds to quasi-particle lifetime(QPLT), is expressed in reciprocal energy units. The contribution $\Sigma_{e,0}$ mentioned above is given by

$$\Sigma_{e,0} = -n_0 \sum_{\mathbf{k}'} |V_0(\mathbf{k}-\mathbf{k}')|^2 \int_{-\infty}^{+\infty} \varepsilon\, d\varepsilon\, \rho_0[\, u_\mathbf{k}^2\,(\omega_n^2 + \varepsilon^2)^{-1} + v_\mathbf{k}^2\,(\omega_n^2 + \varepsilon^2)^{-1}\,] = 0.$$

Upon using the Dyson's equation, the full propagator may be written as

$$G^{(Full)}_{\alpha,\alpha}(\mathbf{k},\omega_n) = u_{r,\mathbf{k}}^2 [i\omega_n - \acute{\varepsilon}_r(\mathbf{k}) + i(1/4\tau_\mathbf{k}^{(u)}) + \mu]^{-1} + v_{r,\mathbf{k}}^2 [i\omega_n + \acute{\varepsilon}_r(\mathbf{k}) + i(1/4\tau_\mathbf{k}^{(l)}) + \mu]^{-1}, \qquad (7)$$

where the renormalized Bogoliubov coherence factors $(u_{r,\mathbf{k}}^2, v_{r,\mathbf{k}}^2)$ are given by

$$u_{r,\mathbf{k}}^2 = (1/2)[1 - \{(V + i/4\tau_\mathbf{k})/(\acute{\varepsilon}_r(\mathbf{k}) + V/(4\tau_\mathbf{k}\acute{\varepsilon}_r(\mathbf{k})))\}],$$

$$v_{r,\mathbf{k}}^2 = (1/2)[1 + \{(V + i/4\tau_\mathbf{k})/(\acute{\varepsilon}_r(\mathbf{k}) + V/(4\tau_\mathbf{k}\acute{\varepsilon}_r(\mathbf{k})))\}],$$

$$V = V(\xi, s_z, a|\mathbf{k}|)) = \{(t'_{so})^2 + (at'_2 |\delta\mathbf{k}|)^2\}^{1/2} + \xi s_z \Delta_z\},$$

$$\acute{\varepsilon}_r(\mathbf{k}) = [\{(a|\mathbf{k}|)^2 + V^2\} - \{1/(16\tau_\mathbf{k}^2)\} + \{1/(16\tau_\mathbf{k}^2((a|\mathbf{k}|/V)^2 + 1))\}]^{1/2},$$

$$(1/4\tau_\mathbf{k}^{(u)}) = (4\tau_\mathbf{k})^{-1} - V/(4\tau_\mathbf{k}\acute{\varepsilon}_r(\mathbf{k})),\ (1/4\tau_\mathbf{k}^{(l)}) = (4\tau_\mathbf{k})^{-1} + V/(4\tau_\mathbf{k}\acute{\varepsilon}_r(\mathbf{k})). \qquad (8)$$

The chemical potential $\mu$, according to the Luttinger rule, is given by the equation 1 = $\int d(\mathbf{k}a) \sum_\nu \rho^{(\nu)}_{Fermi}(\mathbf{k}) \times (\exp(\beta(\acute{\varepsilon}_r^{(\nu)}(\mathbf{k})-\mu))+1)^{-1}$ where $\rho^{(\nu)}_{Fermi}(\mathbf{k})$ is the Fermi energy density of states, $\int d(\mathbf{k}a) \to \int_{-\pi}^{+\pi}(d(k_x a)/2\pi \int_{-\pi}^{+\pi}(d(k_y a)/2\pi$, and $\beta = (k_B T)^{-1}$. Next, we consider the quasi-particle scattering problem within the T-matrix approach[19]. As a necessary step, assuming low concentration of impurities, one may include the contributions of all such diagrams in Fig.2 which involve only one impurity vertex. This gives the equation to determine the total self-energy $\Sigma(\mathbf{k},\omega_n)$ involving vertex function $\Gamma_0(\mathbf{k},-\mathbf{q},\omega_n)$ which in turn is given by the Lippmann-Schwinger equation:

$$\Sigma_e(\mathbf{k},\omega_n) = n_0 \sum_{\mathbf{k}'} V_0(\mathbf{k}-\mathbf{k}') G_{\alpha,\alpha}(\mathbf{k}',\omega_n) V_0(\mathbf{k}'-\mathbf{k})$$

$$+ n_0 \sum_{\mathbf{q},\mathbf{q}',\mathbf{q}''} V_0(\mathbf{q}) G_{\alpha,\alpha}(\mathbf{k}-\mathbf{q},\omega_n) V_0(\mathbf{q}') G_{\alpha,\alpha}(\mathbf{k}-\mathbf{q}-\mathbf{q}',\omega_n) V_0(\mathbf{q}'') \delta(\mathbf{q}+\mathbf{q}'+\mathbf{q}'') + \ldots$$

or,

$$\Sigma_e(\mathbf{k},\omega_n) = n_0 \sum_{\mathbf{k}'} V_0(\mathbf{q}) G_{\alpha,\alpha}(\mathbf{k}-\mathbf{q},\omega_n) \Gamma_0(\mathbf{k},-\mathbf{q},\omega_n), \qquad (9)$$

$$\Gamma_0(\mathbf{k},-\mathbf{q},\omega_n) = V_0(-\mathbf{q}) + \sum_{\mathbf{q}'} V_0(\mathbf{q}'-\mathbf{q}) G_{\alpha,\alpha}(\mathbf{k}-\mathbf{q}',\omega_n) \Gamma_0(\mathbf{k},\mathbf{q}',\omega_n). \qquad (10)$$

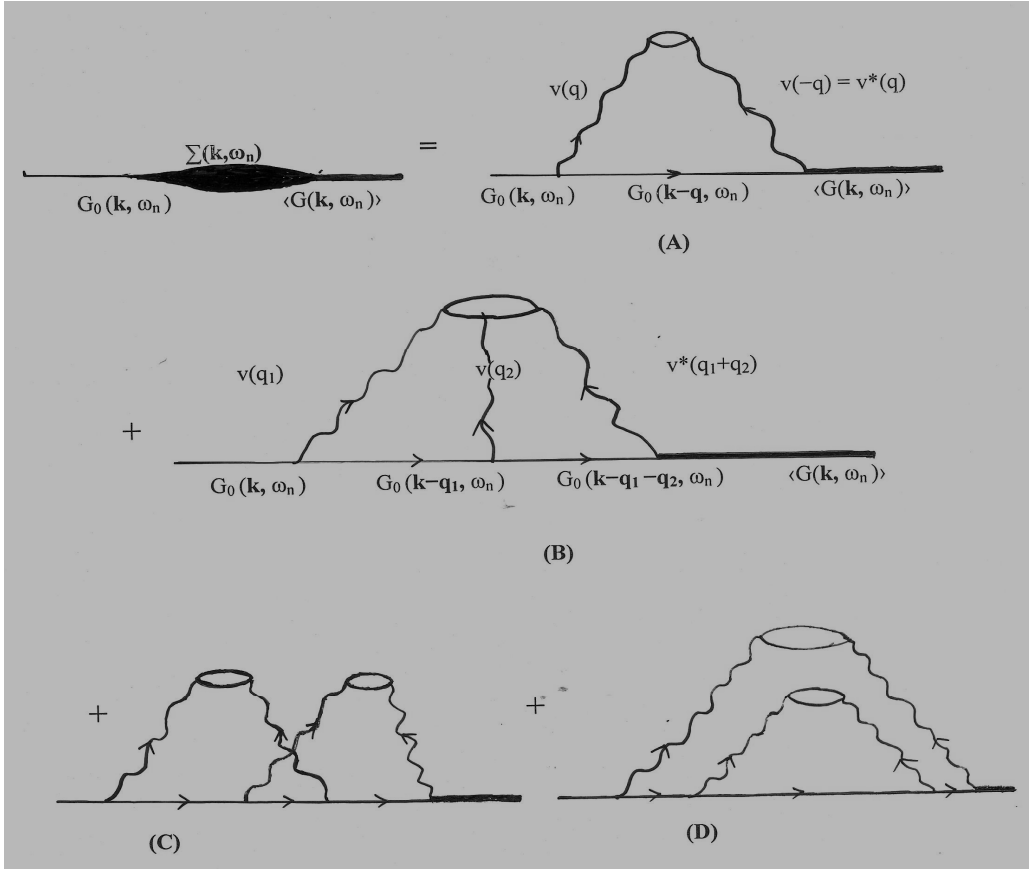

**Figure 2.** A few diagrams contributing to the self-energy. The wiggly lines carry momentum but no energy. The total momentum entering each impurity vertex, depicted by a slim ellipse, is zero. We have assumed that impurities are alike, and distributed randomly. Whereas Figs.(A) and (B) correspond to one impurity vertex, the Figs.(C) and (D) correspond to a product of four impurity potentials with non-zero averages. These are the cases where two impurities each give rise to two potentials. Thus the figures involve the interference of the scattering by more than one impurity. We have assumed low concentration of impurities and therefore these figures yield smaller contributions compared to those corresponding to (A), (B) and the other diagrams of the same class involving only one impurity vertex.

Similarly, the equation for $\Sigma_{so}(\mathbf{k},\omega_n)$ and the one for the corresponding vertex function could be written down. This is the t-martix approximation. Upon using the optical theorem for the T-matrix [20] one may write $\Sigma(\mathbf{k},\omega_n) = i\, \mathrm{Im}\, \Gamma(k,k,\omega_n) = -i\omega_n/(2|\omega_n|\acute{\Gamma}_k)$ where $\acute{\Gamma}_k^{-1} = 2\pi\rho_0 \sum_{\mathbf{k}'}(n_0|\Gamma_0(\mathbf{k},\mathbf{k}')|^2 + n_{so}|\Gamma_{so}(\mathbf{k},\mathbf{k}')|^2)$. Thus the effect of the inclusion of contribution of all the above mentioned diagrams is to replace the Born approximation for scattering by the exact scattering cross-section for a single impurity, i.e. $\tau_k^{-1} \rightarrow \acute{\Gamma}_\mathbf{k}^{-1}$. Since $G_{\alpha,\alpha}(\mathbf{k},\omega_n)$ and $(V_0(\mathbf{k}), V_{so}(\mathbf{k}))$ are known, using Eqs.(9), (10), and Luttinger rule one can determine $\acute{\Gamma}_\mathbf{k}^{-1}$ in terms of $(V_0(\mathbf{k}), V_{so}(\mathbf{k}))$. Upon considering the impurity scattering effect above, we are now in a position to write down the renormalized single-particle excitation spectrum:

$$E_{renorm}^{(U,L)}(\mathbf{k}) = \pm \acute{\varepsilon}_r(\mathbf{k}) - \mu', \qquad (11)$$

$$\acute{\varepsilon}_r(\mathbf{k}) = [\{(a|\mathbf{k}|)^2 + ((t'_{so}{}^2 + (at'_2|\delta\mathbf{k}|)^2)^{1/2} + \xi s_z \Delta_\mathbf{z})^2\} - (1/16\, \acute{\Gamma}_\mathbf{k}^2)\{(a|\mathbf{k}|)^2/(V^2 + (a|\mathbf{k}|)^2)\}]^{1/2}. \qquad (12)$$

The excitation spectrum without the effect of the impurities is given above as $[\pm\{(a|\mathbf{k}|)^2 + ((t'_{so}{}^2 + (at'_2|\delta\mathbf{k}|)^2)^{1/2} + \xi s_z \Delta_\mathbf{z})^2\}^{1/2} - \mu']$. So, as already observed, the effect of impurity scattering is to

alter the excitation spectrum in a fundamental way.

## 4. CONCLUDING REMARKS

Upon treating Eq.(12) as the focal point of the discussion, we notice that, as long as the (non-magnetic) impurity scattering strength is moderate, i.e. the potential strength is of the same order or less than $t_{so}$(~ 4 meV), VSPM phase is protected. The reason being, tuning of $E_z$ allows to arrive at a critical value ($E_c$). In fact, at the critical point with ($E_z/E_c$) = 1.00 the gap of one of the spin-split bands ($s_z = \pm 1$) closes to give a Dirac point while at the other **K** point it is gapped. Furthermore, it is the other spin band which has no gap. For example, for $s_z = -1$ and $\xi = +1$, one has gap closing, i.e. $E_{renorm}$ (**k**, $s_z = -1$, $\xi = +1$) $\approx \pm [\{(a|\mathbf{k}|)^2 + ((t'_{so}{}^2 + (at'_2|\delta\mathbf{k}|)^2)^{1/2} - \Delta_z)^2\}]^{1/2} - \mu'$ $\approx \pm(a|\mathbf{k}|)$ for $\mu' = 0$ due to $\Delta_{zc} = \Delta_z \approx (t'_{so}{}^2 + (at'_2|\delta\mathbf{k}|)^2)^{1/2}$. However, in this case at the other **K** point $E_{renorm}$ (**k**, $s_z = -1$, $\xi = -1$) $\approx \pm[\{(a|\mathbf{k}|)^2 + 4\Delta_z{}^2\}]^{1/2}$ is gapped. One also finds for the other spin band no gap($E_{renorm}$ (**k**, $s_z = +1$, $\xi = -1$) $\approx \pm(a|\mathbf{k}|)$ for $\mu' = 0$) and gap( $E_{renorm}$ (**k**, $s_z = +1$, $\xi = +1$) $\approx \pm[\{(a|\mathbf{k}|)^2 + 4\Delta_z{}^2\}]^{1/2}$).The effective "two-component Dirac physics" thus remains valid in this phase. The increase in $E_z$ beyond the critical value ($E_z/E_c$) = 1.00 leads to the orbital magnetic moment (M) reversal. It must be noted that magnetic moment in silicene has both orbital and spin character. In addition to the spin, the Bloch fermions carry the orbital magnetic moment [21] due to the self-rotation of the wave packets around its centre of mass. Under symmetry operations, the orbital moment transforms exactly like the Berry curvature in silicene [22]. Interestingly, by actual calculation as in [22], it is found to be proportional to the expression of the Berry curvature of the conduction band:

$$M_\xi (\xi, s_z, a\, |\delta\mathbf{k}|) \sim \xi \times [\, V(\xi, s_z, a\, |\delta\mathbf{k}|) / \{(V(\xi, s_z, a\, |\delta\mathbf{k}|))^2 + (a|\delta\mathbf{k}|)^2\}^{3/2}], \qquad (13)$$

$$V(\xi, s_z, a|\delta\mathbf{k}|) = \{\Delta_{SOC}(a\, |\delta\mathbf{k}|) + \xi s_z \Delta_z\}, \quad \Delta_{SOC}(a\, |\delta\mathbf{k}|) = (t'_{so}{}^2 + (at'_2|\delta\mathbf{k}|)^2)^{1/2}, \qquad (14)$$

and $s_z = \pm 1$ for $\{\uparrow, \downarrow\}$. Therefore unless the system has both time-reversal and inversion symmetry, the orbital moment is in general nonzero. Since the valley index $\xi$ determines the sign of the orbital magnetic moment, the latter may also be termed as the valley magnetic moment (VMM). This is estimated to be two times greater than that of graphene[21]. Therefore, an applied magnetic field is expected to elicit greater response from silicene. Naturally, silicene/germanene is a better options to realize valley polarization than graphene. In Figure3, we have plotted the valley magnetic moment as a function of the dimensionless electric field close to the Dirac point. We find that, the critical point with $(E_z/E_c) = 1.00$ is characterized by the VMM (M) sign reversal. The VMM, in fact, vanish everywhere except at the Dirac points where they diverge.

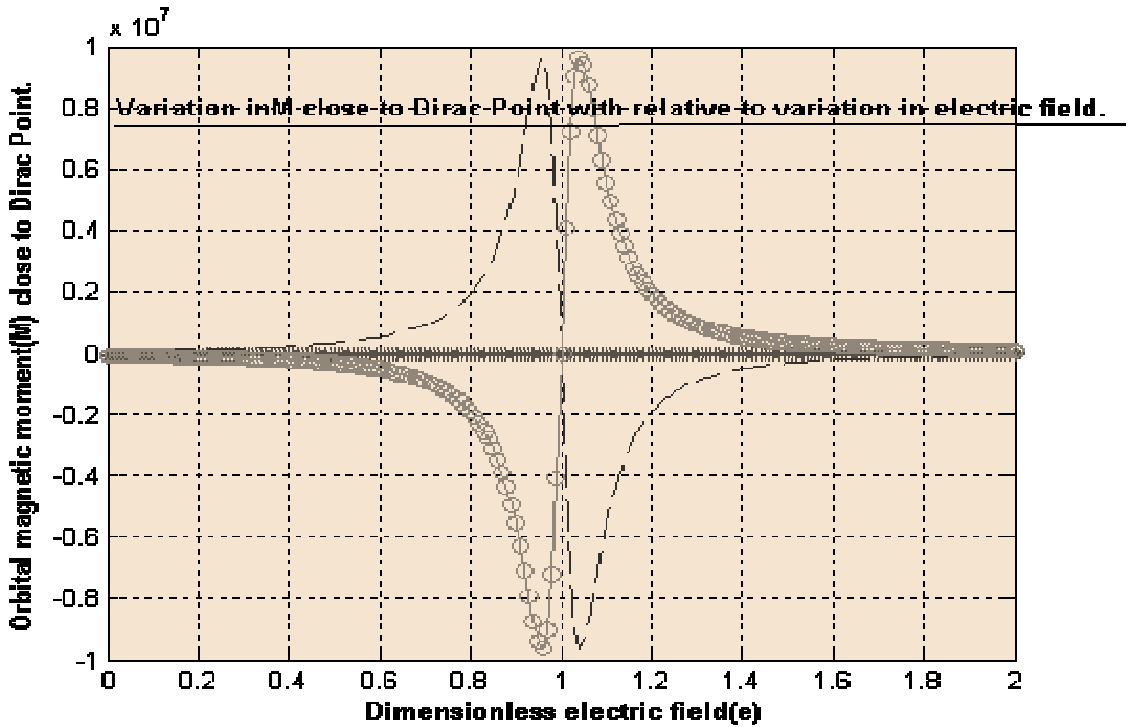

**Figure 3.** In this figure we have plotted the VMM (M)as a function of the dimensionless electric field(e) close to the Dirac point. We find that, the critical point with $(E_z/E_c) = 1.00$ is characterized by the VMM sign reversal.

The increase in the impurity scattering strength for the elastic scattering and the spin-orbit scattering leads to the disappearance of the SVPM phase. One can see this easily, for example, for the case $s_z = -1$ and $\xi = +1$. In this cae one has gap closing, i.e. $E_{renorm}$ (**k,** $s_z = -1, \xi = +1$) ≈ ± ($a|\mathbf{k}|$) for $\mu' = 0$ only when $\Delta_{zc} = \Delta_c \approx [(t'^2_{so} + (at'_2|\delta\mathbf{k}|)^2)^{½} - (\pm 1/4\ \acute{\Gamma}_\mathbf{k}) \{(a|\mathbf{k}|)^2/(V^2 + (a|\mathbf{k}|)^2)\}] \approx -(\pm 1/4\ \acute{\Gamma}_\mathbf{k})$. The last line appears due to the reasons that $\{(\left(\frac{\hbar v_F}{a}\right)a|\mathbf{k}|)^2/(\left(\frac{\hbar v_F}{a}\right)V)^2 + (\left(\frac{\hbar v_F}{a}\right)a|\mathbf{k}|)^2)\} \approx 1$ and $(t'^2_{so} + (at'_2|\delta\mathbf{k}|)^2) \ll (1/16\ \acute{\Gamma}_k^2)$. However, at the other **K** point one has gap: $E_{renorm}$ (**k,** $s_z = -1, \xi = -1$) $\approx \pm[\{(a|\mathbf{k}|)^2 + 4\Delta_z^2\}]^{½}$. Particularly, $\Delta_{zc} = \Delta_z \approx \mp (1/4\ \acute{\Gamma}_\mathbf{k})$ means the 'so-called' gap closing could be accessed only at a unreasonably high value of the applied electric field. The inescapable conclusion is the disappearance of the VSPM phase. With $t_2$ ~ 1 meV (less than $t_{so}$ ~ 4 meV) we do not expect the Rashba coupling to play a major role in the VSPM issue. So, the transition beyond the effective single-valley Dirac Physics is not encouraged by the enhancement in $t_2$.